\documentclass{article}

\usepackage[utf8]{inputenc} 
\usepackage[T1]{fontenc}    
\usepackage{hyperref}       
\usepackage{url}            
\usepackage{booktabs}       
\usepackage{amsfonts}       
\usepackage{nicefrac}       
\usepackage{microtype}      
\usepackage{xcolor}         
\usepackage{graphicx}
\usepackage{subfig}
\usepackage{todonotes}

\definecolor{brilliantrose}{rgb}{1.0, 0.33, 0.64}
\definecolor{myviolet}{rgb}{0.21, 0.0, 0.85}
\definecolor{amethyst}{rgb}{0.6, 0.4, 0.8}
\definecolor{carrotorange}{rgb}{0.93, 0.57, 0.13}

\title{Quantum Deep Learning: Sampling Neural Nets with a Quantum Annealer}

\author{
  Catherine F.~Higham \\
  School of Computing Science\\
  University of Glasgow\\
  Glasgow G12 8QQ \\
  \texttt{Catherine.Higham@glasgow.ac.uk} \\
  \and
  Adrian Bedford \\
  OxBrdgRbtx Ltd \\
  Stratford-upon-Avon CV37 6XU \\
}

\begin{document}

\maketitle
\begin{abstract}

 We demonstrate the feasibility of framing a classically learned deep neural network as an energy based model that can be processed on a one-step quantum annealer in order to exploit fast sampling times. We propose approaches to overcome two hurdles for high resolution image classification on a quantum processing unit (QPU): the required number and binary nature of the model states. With this novel method we successfully transfer a convolutional neural network to the QPU and show the potential for classification speedup of at least one order of magnitude.
\end{abstract}

\section{Motivation}
Deep learning approaches are being applied and refined for classification tasks across many data types (images, video and audio) with high levels of success \cite{BMVC2015_41, 10.1145/3065386, bojarski2016end}. There are many applications where classically trained convolutional neural networks (CNNs) perform very well, sometimes better than human experts \cite{Esteva2017, BUETTIDINH2019e00321, Zhou2021}. However in some circumstances, including security, defence and automated transport, safety-critical benefits would arise if classifications can be computed more quickly. This suggests an important role for quantum computing and we give a proof of principle demonstration that quantum annealing has the potential to address this issue.



The manuscript is organised as follows. In Section \ref{s:bg} we discuss the quantum annealing technology, the underlying quantum physics and the machine learning tool that we implement. We also discuss related work and summarise our main contributions. In Section \ref{s:methods} we give more detail about the implementation and justify our algorithmic choices. Computational results are given in Section \ref{s:results}. We finish in Section \ref{s:disc} with brief discussion.



\section{Background and Related Work}\label{s:bg}
\subsection{Quantum Annealing Technology}
The current D-Wave Advantage System 1.1 contains around 5,000 qubits and 35,000 couplers \cite{dwave}. These qubits can be physically coupled to form networks with real-valued coefficients denoting coupling strength and individual on/off biases. Connectivity per qubit is limited to 15 couplers but can be extended by forming chains of qubits. These coefficients define and constrain relationships between qubits and form a quadratic binary model capable of expressing a range of behaviours. The parameter coefficients of a given model are embedded on the D-Wave network and by posing the problem as energy minimisation, quantum annealing is used to find the low energy states (on/off) of the qubits and hence the most likely formations. 
The ability to rapidly sample from many states and hence characterise the shape of the energy landscape is a key benefit of this technology.

\subsection{Underlying Quantum Physics: The Hamiltonian}
\label{ss:physics}
A classical Hamiltonian gives a mathematical description of some physical system in terms of its energies. 
For most non-convex Hamiltonians, finding the minimum energy state is an NP-hard problem that classical computers cannot solve efficiently. In quantum annealing, the system begins in the lowest-energy state of an initial Hamiltonian, $A$, and as it anneals introduces the problem Hamiltonian, $B$. To do this, a time-like parameter $s$ and annealing functions $A(s)$ and $B(s)$ are introduced such that $A(0) \gg B(0)$ and $B(1) \gg A(1)$. Hence, as the system is annealed, $A(s)$ decreases, $B(s)$ increases and we approach the desired solution states. This approach has the potential for significant benefits in terms of both speed and accuracy, compared with classical computing technology.


For the D-Wave system \cite{dwave}, the Hamiltonian is expressed as follows

\begin{equation}\label{eq:Ham}
    H = \frac{A(s)}{2} \left(\sum_i \hat{\sigma}^{(i)}_x \right)  + \frac{B(s)}{2}
     \left(\sum_i h_i \hat{\sigma}^{(i)}_z  + 
           \sum_i J_{i,j} \hat{\sigma}^{(i)}_z \hat{\sigma}^{(j)}_z \right),
\end{equation}
where $\hat{\sigma}^{(i)}_{x,z}$ are Pauli matrices operating on a qubit $q_i$, and $h_i$ and $J_{i,j}$ are the qubit local field values or biases and coupling strengths or weights, respectively.
In the final state, the qubits take values of either 0 or 1. Hence this provides a classical solution for the problem Hamiltonian $B$ defined by the biases, $h_i$ and the coupling weights, $J_{i,j}$.

\subsection{Machine Learning Approaches: Boltzmann Machines}
\label{ss:ML}
Boltzmann machines \cite{Hinton:2007} are probabilistic models with an energy-based distribution that define a probability for each of the $N$ discrete states in a binary vector given by.
\begin{equation}\label{eq:p}
p(x) = \frac{1}{Z} \exp(-E(x;\theta)).
\end{equation}
Here $E(x;\theta)$ is an energy function parameterised by $\theta$, and $Z =\sum_x \exp(-E(x;\theta))$
is the normalizing coefficient, also known as the partition function, which ensures that $p(x)$ sums to 1 over all the possible states of $x$. The energy function can be represented via a quadratic form $x^TQx$ in which the upper-triangular matrix $Q=Q_{\theta}$ encapsulates the parameters of the quadratic energy function defined by 
\begin{equation}\label{eq:E}
    E(x;\theta) = x^T Q x = \sum_i^N q_{i,i} x_i + \sum_{i < j}^N q_{i,j} x_i x_j,
\end{equation}
where $q_i$ and $q_{i,j}$ are the biases and correlation weights respectively. This expression makes clear the connection between Boltzmann machines, quadratic binary models, and the final Hamiltonian in the second part of equation (\ref{eq:Ham}). This suggests that Boltzmann machines are good candidates for training and evaluation on the D-Wave quantum annealer as the problem can be framed in such a way that the solution to the final Hamiltonian is the low energy state of the problem.
In this work we are interested in transferring classically learned weights to a quantum system for speed up and hence we focus on the evaluation task. 
A restricted Boltzmann machine (RBM) is a special type of Boltzmann machine with a symmetrical bipartite structure \cite{hinton2010}. The set of binary variables is divided into visible (input), $v$, and hidden, $h$, variables. These are analogous to the retina and brain, respectively. 
The hidden variables allow for more complex dependencies among visible variables and are used to learn a stochastic generative model over a set of inputs. All visible variables connect to all hidden variables, but no variables in the same layer are linked. In the classical setting this limited connectivity makes inference and therefore learning easier because analytical expressions can be found for the conditional probabilities 
Gibbs sampling may be used to sample $h$ from $v$ and then $v$ from $h$, leading to an estimated expected value of the model required to update the weights, a process known as contrastive divergence (CD) \cite{hinton2010}. 

\subsection{Related Work}
\label{ss:related}
Finding the optimal parameter values for an artificial neural network requires large amounts of data and many cycles of updates which costs time and energy. Training on a QPU has the potential to produce better models and more quickly than a digital processor. \cite{adachi2015application} investigated estimating the model expectations of RBMs using samples on a 512-qubit D-Wave machine and successfully trained a model with up to 32 visible nodes and 32 hidden nodes per RBM layer. In their tests, they found that this approach achieves comparable or better accuracy with significantly fewer iterations of generative training than conventional CD-based training on a coarse-grained version of the MNIST data set.
\cite{sleeman2020hybrid} also investigated the feasibility of using the D-Wave as a sampler for machine learning. Their work described a hybrid system that combined a classical deep neural network autoencoder with a quantum annealing RBM. 
Their method overcame two key limitations in the 2000-qubit D-Wave processor, namely the limited number of qubits available to accommodate typical problem sizes for fully connected quantum objective functions and samples that are binary pixel representations. Their hybrid autoencoder approach indicated advantage for quantum annealing relative to the use of a classical computer implementation for image-based machine learning and hinted at even more promising results for the next generation D-Wave quantum system. 


In \cite{dixit2020training}, the model expectation of gradient learning for RBM was calculated using a quantum annealer (D-Wave 2000Q), giving much faster results than Markov chain Monte Carlo used in contrastive divergence. 
Most Boltzmann machines use restricted topologies that exclude looping connectivity, as such connectivity creates complex distributions that are difficult to sample. \cite{e22111202} used an open-system quantum annealer to sample from complex distributions and implemented Boltzmann machines with looping connectivity. 

In this work our starting point is a classically trained convolutional neural network (CNN) that maps real valued features to a classification label. Unlike \cite{adachi2015application}, \cite{sleeman2020hybrid} and \cite{dixit2020training}  we do not train on the QPU but investigate transferring the CNN problem to the QPU in order to obtain solutions. However we encounter similar issues: qubit number, binary nature and connectivity. This works looks at how these issues can be addressed on the recent D-Wave Advantage in the context of deep learning transfer.
\subsection{Contributions}
\label{ss:aims}
The main contribution of this work is to show that a trained artificial neural network can be transferred to a quantum computing setting. To do this we use the framework of energy minimisation involving an appropriate quadratic form, where quantum annealing provides samples from the low energy, most likely model states. Results, obtainable in microseconds rather than milliseconds, can be used to estimate a predictive class score. We show how to design a quadratic binary model, the engine of quantum annealing, to behave like a neural network, combining deep learned parameters and layer structures from a classically trained neural network with quadratic binary model parameters. Constraint parameters are adapted so that designated class units act together as a softmax classification unit. We then test this approach on digit image data by finding an appropriate embedding on the D-Wave QPU and transferring the coupling and bias parameters from the classical model to the qubits. 
Key barriers to scaling up to high resolution image/video processing are the binary nature of the variables, the number of qubits that could be assigned to visible units and the limited number of couplers per qubits. We address these issues by showing how real-valued features can be introduced into the system and by introducing novel pooling qubits that reduce the number of connections required. These pooling qubits also add nonlinear behaviour required to model data.
Our approach is outlined in Section \ref{s:methods} and the results of experiments on the D-Wave in Section \ref{s:results}.




\section{Methods: Framing Classification as a Quadratic Binary Model with Constraints}
\label{s:methods}
\subsection{Neural Network Layers}
A typical neural network classification model takes features, $x$, as input and passes these features through a series of mappings and activations leading to a classification score \cite{bengio_dl_book}. These mappings typically include a pooling function, such as max pooling, $\max(x)$, a linear transform function, $x^TW + b$, where $W$ is a matrix 
and $b$ is a vector, and 
softmax activation, $S(x_i) = \frac{e^{x_i}}{\sum_j e^{x_j}}$, to determine a class score. The values of the parameters $W$ and $b$ in each layer are determined through training with a cost function to minimise the class score error. In our work we set $b=0$ to reduce the number of parameters. 
\subsection{Quadratic Binary Model with Constraints}
A quadratic binary model defines an energy-based network of binary random variables with real-valued parameters for biases and correlation weights. Constraints can be added to the biases. 
Energy minimisation involves finding the binary values of the model states that result in the lowest energy levels. This minimisation problem is also referred to as quadratic binary optimisation or QUBO.

Let $v = \{v^{ij}\}_{i=1 \ldots N_p, j=1 \dots M_p}$, $h_p = \{h_p^j\}_{j=1 \ldots M_p}$, and $h_c = \{h_c^k\}_{k=1 \ldots M_c}$ be binary random variables where $v$ denote visible units, $h_p$ and $h_c$ denote hidden pooling and classification units, respectively. $N$, $N_p$, $M_p$, $M_c$ are the number of visible units, the number of visible units in a pool, the number of hidden pooling units and the number of class units, respectively. For pooling, $M_p$ subsets of $v$, each containing four neighbouring elements are connected to a $h_p$ reducing the possible connections from $NM_p$ to $4M_p$. All $h_p$ are connected to all $h_c$ forming a fully connected classification layer. Further, $h_c$ are interconnected so that a decision can be made about the most likely state equivalent to a softmax activation.

The energy of this system involving $v$, $h_p$ and $h_c$ and the connections described above can be 
introduced, as in equation (\ref{eq:E}), as follows 
\begin{equation}\label{eq:ENN}
    E(v,h_p,h_c) = -q_v v - \sum_j^{M_p} \sum_i^{N_p} (h_p^j q_{vp}^{ij} v^{ij} + q_p^j h_p^j) 
    - h_c^T q_{pc} h_p -q_c^T h_c - \sum_{j > k}^{M,M-1} h_c^j q_{cc}^{jk} h_c^k,
\end{equation}
where $q_v$, $q_{hp}$, $q_{hc}$ are the biases of $v$, $h_p$, $h_c$ respectively, and $q_{vp}$, $q_{pc}$, $q_{cc}$ are the coupling strengths between $v$ and $h_p$, $h_p$ and $h_c$ and between $h_c$ and $h_c$ respectively.

In order that the quadratic binary model behaves like the neural network classification model, the parameters, $Q=\{q_v,q_{hp},q_{hc},q_{pv},q_{cp},q_{cc}\}$, are set as follows to achieve pooling and classification.

\subsection{Downsampling with Pooling Units}
The objective of pooling for the hidden units $h_p$ is to establish whether features are present. If features are present, then $h_p$ is on and contributions are made to the classification units $h_c$. If features are not present then $h_p$ is off and no contributions are made to the classification units. Consider the contribution a pooling subset, $h_p^j$ makes to the energy equation, 
\begin{equation}
- \sum_i (q_v^i v^{ij} + h_p^j q_{vp}^{ij} v^{ij} + q_p^j h_p).
\end{equation}

If $q_v^i$ are the normalised feature values of $v$, real values between $-\alpha$ and $\alpha$, then neutralising $h_p$ by setting $q_p^j=0$ and weighting the interaction between $v$ and $h_p$ by setting $q_{vp}=\alpha$ lowers the energy of the system and increases the likelihood that $h_p$ is on if a feature is present.

\subsection{Establishing a Classification Unit}

We assume that the weights of a CNN classification layer, $W$, have been optimised so that the class position of the maximum value of the predicted class vector $y=x^TW$, where $x$ is a feature vector, agrees with the class of $x$. The contribution on the classification units $h_c$ comes from an interaction with $h_p$ ($h_c^T q_{pc} h_pc$) and interactions with other $h_c$ ($q_c^T h_c + \sum_{j > k}^{M,M-1} h_c^j h_c^k$), see equation (\ref{eq:ENN}). Concerning the interaction between $h_p$ and $h_c$, we set $q_{pc}=W$. 

For the interaction of $h_c$ values with each other we constrain the model so that $\sum_k h_c^k =1$. This is achieved by finding the values of $q_c$ and $q_cc$ that minimise $(\sum_{k}^{M_c} h_c^k - 1)^2$ using the following expansion 
\begin{equation}\label{eq:NN_E}
    (\sum_{k}^{M_c} h_c^k - 1)^2 = 2 \sum_{j > k}^{M_c,M_c-1} h_c^j q_cc^{jk} h_c^k  - \sum_{j}^{M_c} h_c^j + 1,
\end{equation}
with solutions $q_{cc}=2$ and $q_{c} = -1$.


\subsection{Transfer to QPU}

\begin{figure}[!tbp]
  \centering
  \subfloat[Structure of $x^T Q x$]{\includegraphics[width=0.4\textwidth]{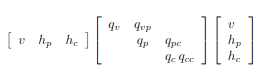}\label{fig:f1}}
  \hfill
  \subfloat[The matrix
  Q]{\includegraphics[width=0.6\textwidth]{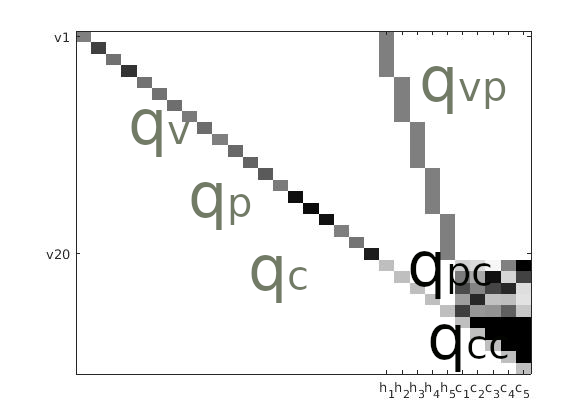}\label{fig:f2}}
  \caption{\textbf{Illustrative Q matrix} for feature pooling and classification. Five sets of four features are each coupled, with weights $q_{vp}$, to a pooling node, see columns labelled $h_1$ to $h_5$. These nodes are further fully connected to five class nodes with coupling strength determined by the learned CNN weights, $q_{pc}$. Further connections are added between the class nodes in order to constrain the class states to just one being on, $q_{cc}$. The biases, $\{q_v, q_p, q_c\}$ associated with each node, including the input features, are placed along the diagonals.\label{fig:Qpool}}
\end{figure}

The weights and biases are physically transferred to the QPU in a sparse matrix format, $Q_{IJ}$, with the biases for each node ${q_v,q_p,q_c}$ along the diagonal in position $Q_{II}$ and the weights connecting each pair ${q_vp,q_pc,q_cc}$ in position $Q_{IJ}$. An illustrative example of $Q$ is shown in Figure \ref{fig:Qpool}. Before the quantum annealing process, a mapping has to be found that physically embeds the $Q$ matrix on the D-Wave graph. This embedding is found using the available heuristic tool, \textbf{minorminer} by \cite{cai2014practical}, at the beginning of the first run and then reused in subsequent runs. 

The D-Wave sampler is called and the returned samples are ordered lowest energy to highest energy and summed across classes to provide a consensus class score and hence classification.


\subsection{Real Valued Features and Scaling Up}

In work, e.g. \cite{adachi2015application, dixit2020training}, where where RBMs are trained on D-Wave the visible units are clamped by adding clamp strength as a constraint on the biases ensuring that $v$ will retain its value in the annealing process. Here, we use the fact that $v$ is known, so the second term in equation (\ref{eq:E}) becomes dependent on $x_i$ only and $q_{i,\hat{j}}$ can be seen as biases. 
In summary, real valued information about the input features entered the system through the biases $q_v$ placed on the visible units $v$. 

\begin{figure}
  \centering
  \fbox{\includegraphics[width=10cm]{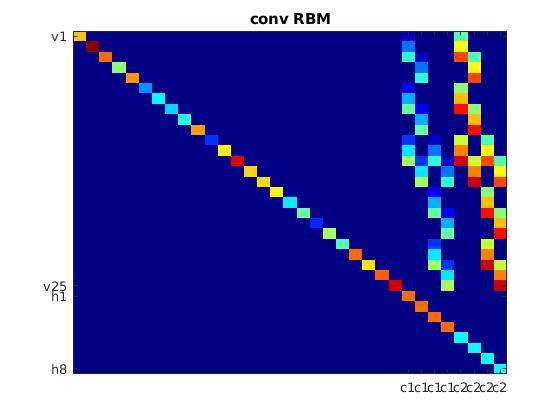}}
  \caption{\textbf{Illustrative Q Matrix for a Convolutional Layer} comprising two convolutional filters, $W_1$ and $W_2$, each containing $3 \times 3$ weights, passing over an image input ($5 \times 5$) with stride 2. This results in two downsized features ($2 \times 2$) where each interaction between input nodes, denoted $v_1 \ldots v_{25}$, and output nodes, denoted $h_1 \ldots h_{8}$, can be represented in matrix form as illustrated above. The interactions include biases (local field values), $b$, along the diagonal, and shared weights, $W_1$ and $W_2$, down the columns $h_1 \ldots h_4$ and $h_5 \ldots h_8$ respectively. Each column denotes one of the eight output features.}
  \label{fig:Qconv}
\end{figure}

Fully connected layers involving large numbers of units are not suitable for the QPU. Instead we investigate the feasibility of using convolutional connections where connections between layers are reduced to a subset of units in each layer but where weights are shared for local and global connectivity. How these weights form part of the energy system $Q$ is illustrated in Figure \ref{fig:Qconv}. 

\section{Results}
\label{s:results}
In this Section we present results from sampling with a quantum annealer for two classification neural networks with five and ten classes, respectively. We finish the Section with a discussion on scaling up and timings.

\subsection{Binary Digits 0:4 Classification}
\begin{figure}
  \centering
  \fbox{\includegraphics[width=10cm]{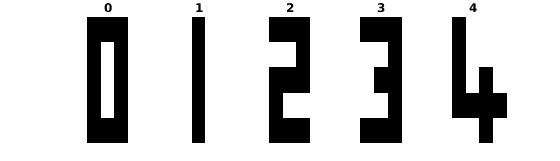}}
  \caption{Binary digits $5 \times 5$.}
  \label{fig:TX}
\end{figure}

A small classification neural network was classically trained to distinguish between binary $5 \times 5$ images for digits $\{0,1,2,3,4\}$ and optimised model parameter values, $W$, were obtained. The network comprised 30 nodes: 20 visible feature nodes, 5 hidden nodes and 5 classification nodes (see Appendix for more details). The Q values for the QUBO were set as follows: $q_pc=W$, $q_vp=1$, $q_{cc}=2$ and $q_c=-1$. Input to this model were binary $5 \times 5$ images for digits $\{0,1,2,3,4\}$, see Figure \ref{fig:TX}, and the features included in the quadratic binary model were extracted from the convolutional layer after sigmoid activation. The biases and weights were then transferred to the QPU and 1000 samples taken. The 100 samples with the lowest energy were averaged to estimate the expected value of each classification node. The results for the five nodes assigned to the five classes clearly show dominance in the correct class, see Table \ref{tbl:bindig}. This simple example shows proof of principle that the weights from a classically trained CNN can be used to connect qubits for classification tasks.

\subsection{MNIST classification}
We investigate using the high level abstracted features from the penultimate layers of a neural network model trained using the MNIST database \cite{lecun-98}. 
The quadratic binary model comprised 50 nodes: 40 feature nodes and 10 classification nodes. The Q values were $q_c=-1$, $q_{cc}=2$ and $q_pc=W$ where $W$ was obtained from the classification layer of the MNIST CNN. The biases and weights were then transferred to the QPU and 1000 samples taken. The 100 samples with the lowest energy were averaged to estimate the expected value of each classification node. The results for the ten nodes assigned to the ten classes show dominance in the correct class for digits 0, 3, 5 and 6. There was confusion between digits 1, 7 and 9. Digits 2 and 4 were never predicted whereas digit 8 was sometimes predicted but with a high error rate, see Table \ref{tbl:bindig}. This experiment was performed on only one randomly chosen digit from each class and with one embedding for Q. With physical experiments, variations are to be expected with the set-up, for example within the individual qubits and couplers. Future work will run this experiment many more times to reduce this variation and we would expect to see more reliable predictions.
This example shows that including high level abstracted features arising from one source or a combination as sources as biases in the the quadratic binary models opens up the classification task to a broader field.

\begin{table}
  \caption{Binary Digit Classification. Class predictions \emph{columns} based on 100 samples with the lowest energy out of 1000 samples are shown for each input class \emph{rows}. All digits show dominance in the correct class.}
  \label{tbl:bindig}
  \centering
  \begin{tabular}{crrrrr}
    \toprule
   \multicolumn{6}{c}{Class Predictions}                   \\
    \cmidrule(r){2-6}
    Input Classes     & 0 & 1 & 2 & 3 & 4 \\
    \midrule
    0 & \textbf{100} & 0  & 4 & 0 & 17     \\
    1     &      0  &  \textbf{100} & 0 & 68 &  0      \\
    2     & 0   &  0 &  \textbf{100}  &  2 &         0  \\
    3     & 0   &      0    &     0  &  \textbf{100}    & 86  \\
    4     & 47 & 0 &   27    & 22   & \textbf{100}  \\
    \bottomrule
  \end{tabular}
\end{table}

\begin{table}
  \caption{MNIST Classification. Class predictions \emph{columns} based on 100 samples with the lowest energy out of 1000 samples are shown for each input class \emph{rows}. Digits 0,1,3,5,6 and 7 show dominance in the correct class. However there is confusion between 1,7 and 9. Digits 2, 4 and 8 are wrongly classified. }
  \label{tbl:MNIST}
  \centering
  \begin{tabular}{crrrrrrrrrr}
    \toprule
   \multicolumn{11}{c}{Class Predictions}                   \\
    \cmidrule(r){2-11}
    Input Classes     & 0 & 1 & 2 & 3 & 4 & 5 & 6 & 7 & 8 & 9\\
    \midrule
    0     & \textbf{98}   & 1   &   29  &   0   &  0    & 0    &17   & 60   &0    & 0     \\
    1     & 10   & \textbf{100} &   0   &   0   &  0    & 0    & 0   & 100  &0    & 0     \\
    2     & 36   & 100 &   0   &   0   &  0    & 0    & 0   & 100  &0    & 0  \\
    3     & 0    & 14  &   1   &   \textbf{86}  &  0    & 79   & 2   & 29   &0    & 0  \\
    4     & 55   & 1   &   2   &   12  &  0    & 14   & 20  & 1    &31   & 46  \\
    5     & 0    & 0   &   0   &   69  &  0    & \textbf{100}  & 40  & 0    &0    & 0 \\
    6     & 88   & 0   &   0   &   0   &  0    & 100  & \textbf{100} & 0    &0    & 0 \\
    7     & 0    & 100 &   0   &   0   &  0    & 0    & 0   & \textbf{100}  &0    & 0 \\
    8     & 96   & 38  &   0   &   0   &  0    & 0    & 0   & 50   &42   & 0 \\
    9     &  0   & 100 &   0   &   0   &  0    & 0    & 0   & 100  &0    & 0 \\
    \bottomrule
  \end{tabular}
\end{table}



\subsection{Scaling Up and Timing}

Using our approach we were able to find an embedding and obtain samples from a four layer model comprising 358 nodes: 196 visible nodes, 128 convolutional feature nodes, 24 fully connected hidden nodes and 10 classification nodes. The convolutional window was $5 \times 5$ and eight windows passed over the input in strides of two, resulting in $8 \times 16$ outputs. We repeated the experiment on a classical computer and estimated that classification and printing out would take about 8000 microseconds whereas samples can be obtained from the QPU in 168 microseconds (anneal 20 µs Readout 127 µs Delay 21 µs) suggesting a possible speed up of at least one order of magnitude.

\section{Discussion and Conclusions}\label{s:disc}

We have shown that it is possible to frame a classification task as a quadratic binary model using the weights from a classically trained classification neural network. This problem can then be sent to D-Wave's QPU for one-step quantum annealing. Given a trained network, D-Wave can be used to evaluate it.

Regarding timings, we have focused on the annealing time (around 20 $\mu s$). Currently, there are time overheads (access, programming, sampling and post-processing) involved with this process but these could be addressed by engineering and pipelines that improve streaming to and from the QPU for specific tasks such as classification.

We addressed two barriers to scaling up: binary input states which restrict types of data that can be analysed and number of couplings between nodes limited by the physical D-Wave graph; and required number of the model variables. We showed that real value features can be considered as biases that can be added directly to the biases of connecting nodes or subject to pooling and connecting node constraints. This illustration opens up the possibility of different types of input features including features derived from other systems or multi-faceted features from complex systems. We introduced pooling and convolutional couplings. Pooling serves two main purposes. First, down-sampling the input which can reduce the size of the network and hence the number of parameters needed (especially important for quantum computers), and reduce the computational cost, and can improve generalisation. Second, adding non-linearity, required for better model expressiveness, to the linear maps. 
Convolutional filters also reduce the connectivity load, and by extracting features from spatial settings improve model expressiveness. The largest network we were able run had four layers and 358 nodes.

In summary, providing neural networks with a quantum engine has the potential, assuming the pipeline for streaming data to the quantum computer can be made more efficient, to obtain classification results from high dimensional sources with speeds at least an order of magnitude.

\clearpage
\bibliography{references}
\bibliographystyle{plain}

\section{Appendix}

\subsection{Classically Trained Neural Network Specifications}
\subsubsection{Binary 5 x 5 Digits 0:4 Classification}
The classically trained convolutional neural network comprised input, convolutional (filter size $W=3$, $H=3$ and $D=5$, stride=2) sigmoid activation, max pooling (window $W=2$ and $H=2$, stride=2), fully connected (filter size $W=5$ and $H=5$, and softmax activation layers where $W$ denotes width and $H$ height. For input sized $5 \times 5$ the resulting features after the convolutional layer are sized $2 \times 2 \times 5 $ and after the pooling layer are sized $1 \times 5$.

\end{document}